\documentclass[a4paper,11pt]{article}

\usepackage{bm}
\usepackage{amssymb}
\usepackage{amsmath}
\usepackage{graphicx}
\usepackage{subfigure}
\usepackage{multirow}
\usepackage{authblk}
\usepackage{fullpage}

\newcommand{\Keywords}[1]{\par\noindent
{\small{\em Keywords\/}: #1}}

\title{Monotone Splines Lasso}
\author{Linn Cecilie Bergersen, Kukatharmini Tharmaratnam and Ingrid K. Glad}
\affil{Department of Mathematics, University of Oslo}
\date{}

\usepackage[authoryear,comma]{natbib} 
\usepackage{bibentry}
\bibpunct{(}{)}{,}{a}{,}{,}
\begin{document}
\maketitle
\normalsize

\abstract{We consider the problems of variable selection and estimation in nonparametric additive regression models for high-dimensional data. In recent years, several methods have been proposed to model nonlinear relationships when the number of covariates exceeds the number of observations by using spline basis functions and group penalties. Nonlinear {\it monotone} effects on the response play a central role in many situations, in particular in medicine and biology. We construct the monotone splines lasso (MS-lasso) to select variables and estimate effects using monotone spline (I-splines). The additive components in the model are represented by their I-spline basis function expansion and the component selection becomes that of selecting the groups of coefficients in the I-spline basis function expansion. We use a recent procedure, called cooperative lasso, to select sign-coherent groups, that is selecting the groups with either exclusively non-negative or non-positive coefficients. This leads to the selection of important covariates that have nonlinear monotone increasing or decreasing effect on the response. We also introduce an adaptive version of the MS-lasso which reduces both the bias and the number of false positive selections considerably. We compare the MS-lasso and the adaptive MS-lasso with other existing methods for variable selection in high dimensions by simulation and illustrate the method on two relevant genomic data sets. Results indicate that the (adaptive) MS-lasso has excellent properties compared to the other methods both by means of estimation and selection, and can be recommended for high-dimensional monotone regression.
\vspace{2mm}
}

\section{Introduction}\label{s:intro}
In certain bio-medical applications it is important to assume that the relationship between an explanatory variable and the outcome is monotonically increasing or decreasing. Actually, every time linear regression is applied, an implicit monotonicity assumption is made. Monotone, but not necessarily linear relations appear typically for dose-response data, for example. It is also reasonable to assume that the relation between a disease and a risk factor is monotone, but not necessarily linear \citep{Sylvia}.

There has been a major effort in developing methods for monotone regression beyond the strictly linear regression models. These methods are usually concerned with classical situations in which the number of covariates $P$ does not exceed the number of observations $n$. In the last
decade, the massive production of data sets in all areas of science and technology has turned high-dimensional regression problems, where $P$ is much larger than $n$, into one of the most active research areas. Very recently, an important contribution has appeared for monotone regression in high dimensions \citep{LISO}. In this paper we develop another substantially different tool for this purpose.

When $P$ is larger than $n$, penalized regression methods handle the dimensionality problem by adding a penalty to the log-likelihood to be maximized. The lasso \citep{Lasso} and its many variants (\citet{adaptivLas}; \citet*{adapHighDim2}; \citet{GroupLas1, Elnet}; \citet{ RelLas}) also have the advantage of setting some of the regression coefficients to zero, thus producing a sparse solution. These standard penalized regression methods limit the search to linear relationships between the covariates and the response. Recently nonparametric methods for high-dimensional regression have started to emerge. Recent papers (\citet*{Buhlmann2009, Huang2010}; \citet{SPAM}) consider
a generalized additive model (GAM) \citep{GAM} in combination with spline approximations. Given the observations $(y_i, \bm{x}_i), i=1,...,n$, where $y_i$ is the response and $\bm{x}_i = (x_{i1}, ..., x_{iP})^t$ is the vector of covariates for observation $i$, the additive model is given by \begin{equation} y_i = \beta_0
+ \sum_{j=1}^P g^0_j(x_{ij})+\epsilon_i.\label{gam} 
\end{equation}
Here $\beta_0$ is the intercept, the $g^0_j$'s are unknown functions to be estimated and $\epsilon_i$ is the unobserved independent random error with mean zero and variance $\sigma^2$. We assume ${E}{g^0_j}(\bm{x}_j)=\bm{0},$ for $1 \leq j \leq P$, where now $\bm{x}_j=(x_{1j},\ldots,x_{nj})^t$, to ensure unique identification of the $g^0_j$'s. In \citet{Buhlmann2009, Buhlmann2011, Huang2010} and \citet{SPAM} each nonparametric component $g^0_j$ is represented by a linear combination of spline basis functions and the problem can be viewed in terms of a group lasso problem \citep{GroupLas1}; hence selecting groups of spline basis functions representing relevant covariates. B-splines are popular for representing the covariates, because of their flexibility and minimal assumptions on the form of the function to be estimated. Combined with the group lasso, the framework becomes a highly flexible alternative to (standard) linear lasso-type methods. Our aim is to construct a new method that is nonparametric and flexible as above, but restricted to select and estimate monotone functions $g^0_j$ only. 

In simple regression problems, monotone increasing relationships are often modelled through isotonic regression (\citet{Barlow1972}; \citet{Robertson1988}) which is a nonparametric modelling technique restricting the regression function to be nondecreasing. Additive isotonic models assuming that each component effect in the additive model is isotonic, was presented in \citet{Bacchetti1989}. However, most literature on monotone and isotonic regression is limited to low dimensions. As mentioned, \citet{LISO} recently proposed Lasso Isotone (LISO) which combines estimation of nonparametric isotonic functions with ideas from sparse high-dimensional regression in an additive isotonic regression model, and is, to our knowledge, the only method feasible for monotone high-dimensional problems. In addition, using an adaptive liso approach, \citet{LISO} also present a way of fitting the model without assuming that all effects are either increasing or decreasing, that is, allowing for component effects of different sign. Isotonic regression is probably the best known method for preserving monotonicity, but has the disadvantage of producing step functions, which often has little biological plausibility, instead of smooth functions. For simple regression, it is possible to use an additional smoothing procedure in a second step to obtain a smooth function \citep{HeShi1998}. \citet{Tibshirani2011} also proposed nearly-isotonic regression which involves a penalty term controlling the level of monotonicity compared to the goodness of fit. 

Another way of preserving monotonicity is to fit a smooth monotone function via monotone regression splines \citep{Ramsay1988, HeShi1998}. While \citet{HeShi1998} proposed monotone B-spline smoothing based on a constrained least absolute deviation principle, \citet{Ramsay1988} introduced integrated splines (I-splines), which essentially are integrated versions of M-splines that in combination with strictly positive coefficients will produce monotone increasing smooth functions. I-splines have previously been used in connection with a boosting technique to do monotonic regression in a multivariate model in \citet{Tutz2007}. Also \citet{Meyer2008} considers shape-restricted regression splines through I-splines, but only in the one-dimensional case. 

In this paper we propose a new approach to fit nonparametric additive models under the assumption that each component effect $g^0_j(x)$ is monotone. The \textit{monotone splines lasso} (MS-lasso) combines the idea of I-splines with the cooperative lasso \citep{coop}, and is feasible in high-dimensional settings where the
number of covariates $P$ can exceed the number of observations $n$. The cooperative lasso is a lasso method where known groups of covariates are treated together, but differs from the standard group lasso \citep{GroupLas1} in that it assumes that the groups are sign-coherent. That is, the covariates inside a group are cooperating, so either the linear coefficients are all nonpositive, all nonnegative or all null inside a group. This can be combined with the idea of monotone I-splines by letting each covariate, represented via an I-splines basis, constitute a group in the cooperative lasso. Thus MS-lasso fits the additive nonparametric regression model with components that can be either nondecreasing, nonincreasing or of no effect. The important advantages of the MS-lasso are indeed that it is not restricted to only monotone
\textit{increasing} effects, that is, the monotone functions $g^0_j$ can be either monotone \textit{increasing} or \textit{decreasing} in the same model, and also it is fitting \textit{smooth} monotone functions to each $g^0_j$. In this way we are more flexible than the linear model, but more restrictive than the pure
nonlinear methods without any shape constraints. Our method is also biologically more relevant than the adaptive liso, in that we obtain smooth representations of the functions right away. We also suggest a two-step estimator, the adaptive MS-lasso, which leads to less bias and fewer false positives in the final model. 

The paper is organized as follows. In Section \ref{sec:mono} we present the MS-lasso and discuss some of its properties. Also the adaptive MS-lasso is presented, and we
discuss connections to related methods. Section \ref{sec:sim} is dedicated to simulation studies. In Section \ref{sec:appl} we illustrate the use of our method in genomic data, before a final discussion is given in Section \ref{sec:disc}.

\section{Monotone Splines Lasso}\label{sec:mono}
Suppose that each of the regression functions $g^0_j$ in the additive model in (\ref{gam}) can be approximated by $g_j$, a linear combination  of $m$ spline basis functions, that is,
\begin{equation} g_j(x) = \sum_{k=1}^m \beta_{jk} \phi_k(x), \quad 1 \leq j \leq P.\label{eq:compfunc} \end{equation}
Here $\phi_k(\cdot)$ is the basis function and $\beta_{jk}$ is the $k^{th}$ spline coefficient for the $j^{th}$ covariate. Note that for the standard linear regression model $m=1$ and $\phi_1(x)= x$.
 
The representation in \eqref{eq:compfunc} is used in \citet{Huang2010} with a cubic B-spline basis and six interior knots. \citet{Huang2010} combine this spline expansion with an (adaptive) group lasso penalty where the group structure is given by the basis functions for each covariate. This construction allows to perform selection of groups corresponding to important covariates. 

Let $\bm{\beta}=(\bm{\beta}_1,\ldots,\bm{\beta}_P)^t \in \mathbb{R}^{Pm}$  with $\bm{\beta}_j=(\beta_{j1},\ldots,\beta_{jm})^t \in \mathbb{R}^m$ for the $j^{th}$ covariate when using $m$ basis functions. The group lasso norm is 
$$||\bm{\beta}||_{group}=\sum_{j=1}^P w_j ||\bm{\beta}_{j}||$$
where $||\cdot||$ is the standard Euclidian norm and $w_j>0$ are fixed weights for each covariate used to adapt the amount of penalty for each group. Typically $w_j$ is the square root of the group size, to take care of situations with very different group sizes. When all covariates are represented by the same number of basis functions $m$, $w_j = 1, \forall j$. 

In the following construction of the MS-lasso we will introduce monotone splines in \eqref{eq:compfunc} and a special penalty, that guarantees monotonicity, in place of the group penalty above. 

\subsection{The MS-Lasso}
For the MS-lasso, suppose without loss of generality that each $x$ is transformed to take values in $[0,1]$. For regression splines of order $l$, we define knots $0 = t_1 = ... = t_{l} < ... < t_{K+l+1} = ... = t_{K+2l} = 1$, where $K$ is the fixed number of interior knots. Following \citet{Ramsay1988} and \citet{Meyer2008} we first define M-splines by $m=K+l$ M-splines basis functions. For order $l=1$ 
\begin{equation*}
 M_k^{(1)}(x) =
\left\{
	\begin{array}{ll}
		\frac{1}{t_{k+1}-t_k}, \quad t_k \leq x \leq t_{k+1}, \\
		  0 \quad \text{otherwise,}
	\end{array}
\right.
\end{equation*}
for $k=1, ..., K+1$. For higher order $l$, the M-splines can be found by recursion from the lower order functions;

\begin{equation*}
 M_k^{(l)}(x) =
\left\{
	\begin{array}{ll}
		  \frac{l[(x-t_k)M_k^{(l-1)}(x) + (t_{k+l} - x) M_{k+1}^{(l-1)}(x)]}{(l-1)(t_{k+l}-t_k)}, \quad t_k \leq x \leq t_{k+l}, \\
		    0 \quad \text{otherwise.} 
	\end{array}
\right.
\end{equation*}
Finally \citet{Ramsay1988} defines the integrated spline (I-spline) basis consisting of monotone functions by integrating 
\begin{equation*}
 I_k^{(l)}(x) = \int_{t_1}^x M_k^{(l)}(u)du, 
\end{equation*}
for $x \in [0,1]$ and $k = 1, ..., K+l = m$. Based on this, for the MS-lasso we write (\ref{eq:compfunc}) as 
\begin{equation} g_j(x) = \sum_{k=1}^m \beta_{jk} I^{(l)}_k(x), \quad 1 \leq j \leq P.\label{eq:compfunc2}\end{equation}

As each spline function $I_k^{(l)}(x)$ is monotone, monotonicity in $g_j(x)$ is achieved by restricting the spline coefficients to be of the same sign. An I-spline with all $\beta_{jk} \geq 0, \forall k$ will produce a monotone nondecreasing function $g_j(x)$, while an I-spline with $\beta_{jk}\leq 0, \forall k$ will produce a nonincreasing function. The cooperative lasso \citep{coop} is designed especially for problems of estimation and selection of parameters in high dimensions where the group structure is known and sign-coherence is assumed within the group. 

We are centering the basis functions to satisfy ${E}{g_j}(\bm{x}_j)=0$, for $1 \leq j \leq P$. Let $z_{ijk}=I_k(x_{ij})-\bar{I}_{jk}$ be the centered I-spline basis function, where $\bar{I}_{jk}= \frac{1}{n} \sum_{i=1}^n I_k(x_{ij})$. Let $\bm{Z}=(\bm{Z}_1, ...,\bm{Z}_P)$ be the $n \times (Pm)$ design matrix where all covariates are represented by a centered I-spline basis and $\bm{Z}_j$ is the $n \times m$ matrix for the $j$th covariate. We use the centered response vector $\bm{y}$ of length $n$. Then the MS-lasso estimates of $\bm{\beta}$ are defined by minimizing the objective function with respect to $\bm{\beta}$, giving
\begin{equation}\label{beta-mono}
\hat{\bm{\beta}}^{MS} = \underset{\bm{\beta} \in \mathbb{R}^{Pm}}{\arg\!\min} \left\{\frac{1}{2} ||\bm{y}-\bm{Z} \bm{\beta} ||^2 + \lambda ||\bm{\beta}||_{coop} \right\},
\end{equation}
where $\lambda \ge 0$ decides the amount of shrinkage and is common to all groups. Here
\begin{equation}||\bm{\beta}||_{coop} = ||\bm{\beta}^+||_{group}+||\bm{\beta}^-||_{group} = \sum_{j=1}^P w_j \left(||\bm{\beta}^+_{j}||+||\bm{\beta}^-_{j}||\right) \label{eq:coop} \end{equation}
is the cooperative lasso norm with $\bm{\beta}^+=(\bm{\beta}_1^+, ..., \bm{\beta}_P^+)^t$ and $\bm{\beta}^- =(\bm{\beta}_1^-, ..., \bm{\beta}_P^-)^t$, where $\bm{\beta}^+_j=(\beta^+_{j1},\ldots,\beta^+_{jm})^t$ and $\bm{\beta}^{-}_j=(\beta^-_{j1},\ldots,\beta^-_{jm})^t$. Hence $\bm{\beta}^+$ and $\bm{\beta}^-$ are the positive and negative parts of $\bm{\beta}$, that is, ${\beta}_{jk}^+=\max(0,\beta_{jk})$ and ${\beta}_{jk}^-=\max(0,-\beta_{jk})$ respectively.

For simplicity we will use the I-splines basis of order $l=2$ which has the closed form \citep{Tutz2007}
\begin{equation*}I_k^{(2)}(x) =
\left\{
	\begin{array}{ll}
		0,  & x \leq t_k, \\
		\frac{(x-t_k)^2}{(t_{k+1} -t_k)(t_{k+2}-t_k)}, &  t_k < x \leq t_{k+1}, \\
& \\
		1-\frac{(t_{k+2}-x)^2}{(t_{k+2}-t_k)(t_{k+2}-t_{k+1})}, & t_{k+1} < x \leq t_{k+2}, \\
	      1, & x > t_{k+2}, \\
	\end{array}
\right.
\end{equation*} 
such that $g_j(x)$ is defined by letting $l=2$ in \eqref{eq:compfunc2}. Finally the estimated monotone effects can be found as 
$$ \hat{g}_j(x) = \sum_{k=1}^m \hat{\beta}^{MS}_{jk}I^{(2)}_k(x), \quad 1 \leq j \leq P.$$
The estimates $\hat{\bm{\beta}}^{MS}$ can be obtained by using the $n \times (Pm)$ design matrix $\bm{Z}$ of centered I-spline basis function expansions in the cooperative lasso algorithm through the R-package \texttt{scoop} available at {\tt{http://stat.genopole.cnrs.fr/logiciels/scoop}}.

\subsection{Properties}
In the following, we assume that each of the true functions $g^0_j$ can be represented as a monotone I-spline function $g_j$ as in \eqref{eq:compfunc2} with $m$ basis functions. To study the properties of the MS-lasso, we need some standard assumptions on the joint distribution of $(\bm{Z}, \bm{Y})$, which are required to guarantee the convergence of empirical covariances. We assume that the columns of $\bm{Z}$ are centered (hence we have zero mean random variables and ${E}[\bm{Z}\bm{Z}^t]$ is the covariance matrix of $\bm{Z}$), that $\bm{Z}$ and $\bm{Y}$ have finite fourth order moments, and that the covariance matrix of $\bm{Z}$ is invertible. Additionally, we need a  specific assumption for the cooperative lasso, namely that in the sign-incoherent groups, all coefficients are nonzero. Furthermore,  we need two suitable variations of the  irrepresentable condition for the cooperative lasso. All of these assumptions are given in detail in (A1)-(A5) in Section 3 in \citet{coop}. Under these assumptions, the consistency of the cooperative lasso is assured for fixed $P$. All required assumptions are also satisfied in our setting to achieve consistency for the MS-lasso for fixed $P$ and fixed number of spline basis functions $m$. We can re-write the optimization problem in \eqref{beta-mono} in the following normalized form,
\begin{equation}\label{beta-mono2}
 \hat{\bm{\beta}}_n^{MS} = \underset{\bm{\beta} \in \mathbb{R}^{Pm}}{\arg\!\min} \left\{\frac{1}{2n} ||\bm{y}-\bm{Z} \bm{\beta} ||^2 + \lambda_n ||\bm{\beta}||_{coop} \right\},
\end{equation}
where $\lambda_n=\lambda/n$.
Using the results in Theorem 2 in \citet{coop} for fixed $P$, the MS-lasso estimator $\hat{\bm{\beta}}_n^{MS}$ is asymptotically unbiased and has the property of exact support recovery,
\begin{equation} \hat{\bm{\beta}}_n^{MS} \xrightarrow{P} \bm{\beta} \hspace{1 cm}\text{and} \hspace{1 cm} \mathbb{P}\left(\mathcal{S}(\hat{\bm{\beta}}_n^{MS})=\mathcal{S}\right) \rightarrow 1, \label{theory} \end{equation}
for every sequence $\lambda_n$ such that $\lambda_n =\lambda_0 n^{-\gamma}, \gamma \in (0,1/2)$ and where $\mathcal{S}=\{j, \bm{\beta}_j \neq \bm{0}\}$. It follows directly from \eqref{theory} that $\hat{g}_j\xrightarrow{P} g_j$. Under the assumption that $g^0_j$ is in the space of monotone I-spline functions with $m$ basis functions, we have that
$$\hat{g}_j\xrightarrow{P} g^0_j \hspace{1 cm}\text{and} \hspace{1 cm} \mathbb{P}\left(\mathcal{S}^*(\hat{g}_j)=\mathcal{S}^*\right) \rightarrow 1,$$ 
where $\mathcal{S}^*=\{j, g^0_j \neq 0\}$. That is, the difference between the estimated spline function $\widehat g_j$  and the true spline function $g^0_{j}$ for each covariate in model \eqref{gam} converges to zero in probability, and the nonzero components are selected correctly with probability converging to one. 

\subsection{The Adaptive MS-Lasso}
Often, a second adaptive step is introduced for the lasso \citep{adaptivLas} and the group lasso \citep{Buhlmann2011} procedures. This involves replacing the penalties by a re-weighted version based on an initial estimator of the regression coefficients. In the high-dimensional case, the lasso or group lasso themselves can be used to obtain the initial estimator. This idea can also be applied to obtain the adaptive MS-lasso by using the penalty in \eqref{eq:coop} and letting 
\begin{equation}\label{adbeta-mono}
\hat{\bm{\beta}}^{AMS} = \underset{\bm{\beta} \in \mathbb{R}^{Pm}}{\arg\!\min} \left\{\frac{1}{2} ||\bm{y}-\bm{Z} \bm{\beta} ||^2 + \lambda \sum_{j=1}^P w_j \left(||\bm{\beta}^+_{j}||+||\bm{\beta}^-_{j}||\right) \right\},
\end{equation}
$$w_j = \left\{
\begin{array}{ll}
 ||\hat{\bm{\beta}}^{MS}_j||^{-1}, \quad & \text{if } ||\hat{\bm{\beta}}^{MS}_j||>0 \\
  \infty, \quad & \text{if } ||\hat{\bm{\beta}}^{MS}_j|| = 0.
\end{array}
\right.
$$
This assumes that the covariates are represented with the same number of basis functions. 

The intuitive idea and motivation are the same as for the adaptive lasso and the adaptive group lasso. For example, the adaptive (group) lasso procedures have the property that the solution is at least as sparse as the initial estimator, and can therefore be used to reduce the number of false positives compared to the standard initial procedures. The adaptive procedures also penalize less for components with large initial estimators, implying less biased estimates than for the standard procedures.

\subsection{Other Selection Procedures}
We compare the performance of the MS-lasso and the adaptive MS-lasso with four other methods which are doing variable selection in high-dimensional regression. Although all of them are suitable in the high-dimensional setting, comparison is not completely fair as the scopes of the various methods are very different. That is, the underlying assumptions on the component functions are different and will lead to different selections depending on what method we use.  It is, however, important to have a clear vision of what are their differences and when each of them should be applied depending on the problem. 

Table \ref{tab:selections} shows an overview of the different methods and the corresponding basis functions and penalties that are used. In Table \ref{tab:selections}, MS-lasso and AMS-lasso are the proposed methods, Lasso and Ad. lasso are standard and adaptive lasso respectively, while Ad. liso is the adaptive LISO in \citet{LISO}, and BS-lasso is the adaptive group lasso using B-splines in \citet{Huang2010}. It is the combination of type of penalty and representation of data that defines the assumptions we make on the component effects. For example, the lasso and the adaptive lasso together with the original data measurements, will be restricted to important linear effects, while the BS-lasso also allows for components describing the response nonlinearly without any restrictions on the shapes in the estimated nonzero effects. \citet{LISO} proposed the adaptive liso for additive isotonic regression models. The adaptive liso considers a two stage procedure, first calculating the initial fit using the liso algorithm. In the second stage, they conduct a liso procedure again with covariate weights based on the initial fit of the model. Finally, the MS-lasso and the adaptive MS-lasso select important monotone, but possibly nonlinear effects with the combination of monotone I-splines and the cooperative lasso penalty. 

\begin{table}
\centering
\caption{Overview of selection procedures.}
\small
\begin{tabular}{lccc}
\hline
Method & Penalty & Basis & Selection \\ 
\hline 
&& & \\
MS-lasso & $\lambda \sum_{j=1}^P \left(||\bm{\beta}^+_{j}||+||\bm{\beta}^-_{j}||\right)$ & Monotone & Nonlinear \\
&& I-splines basis, & monotone \\
AMS-lasso & $\lambda \sum_{j=1}^P w_j \left(||\bm{\beta}^+_{j}||+||\bm{\beta}^-_{j}||\right)$ & $I_k^{(2)}({x})$ & effects\\
&&& \\ 
\hline
&&& \\
\multirow{2}{*}{Ad. liso} & \multirow{2}{*}{$\lambda \sum_{j=1}^P \Delta (g_j)$}& Step function& Monotone effects \\
 & &basis&  (not smooth) \\
&& & \\
\hline 
&&& \\
Lasso & $\lambda \sum_{j=1}^P |\beta_j|$ & \multirow{3}{*}{Original data, $\bm{X}$} & \multirow{3}{*}{Linear effects} \\
&&&\\
Ad. lasso & $\lambda \sum_{j=1}^P w_j |\beta_j|$ & & \\
&&& \\
\hline
&&& \\
\multirow{2}{*}{BS-lasso}  & \multirow{2}{*}{$\lambda \sum_{j=1}^P w_j ||\bm{\beta}_{j}||$} &B-splines basis& \multirow{2}{*}{Nonlinear effects}\\
      & & $B_k^{(2)}({x})$& \\
&&& \\
\hline 
\end{tabular}
\label{tab:selections}
\end{table}

\subsection{Choice of Tuning Parameter}
The MS-lasso and the other methods listed in Table \ref{tab:selections}, all require selection of the penalty parameter $\lambda$ that controls the amount of shrinkage and hence the number of variables selected. To choose $\lambda$ we use K-fold cross-validation on a grid of $\lambda$ values defining the set of candidate models. K-fold cross-validation, typically with $K=10$, involves splitting the data into $K$ folds. Leaving one fold out at a time, the candidate model is fitted on the remaining $9/10$ of the data while predicting for the left out fold. We choose the model minimizing an estimate of the prediction mean squared error,
$$CV(\lambda) = \frac{1}{n} \sum_{k=1}^K \sum_{i \in f_k} (y_i-\hat{y}_i^{-k}(\lambda))^2,$$ where $f_k$ is the set of indices of the samples in fold $k$, and $\hat{y}_i^{-k}(\lambda)$ is the fitted predicted value for observation $i$ when fold $k$ involving observation $i$ is left out of the estimation. For the adaptive approaches, both the initial estimator and the final adaptive estimator can be estimated with $\lambda$ chosen by cross-validation. 

\section{Simulation Studies}
\label{sec:sim}
To investigate the finite-sample performance of our method, we report the results from several simulation experiments. In all experiments we use the MS-lasso and the adaptive MS-lasso to estimate the component effects $g_j$ and compare with the methods listed in Table \ref{tab:selections}. For the monotone splines methods we use a monotone I-splines basis of order two and with six evenly distributed knots for all functions $g_j$. For the BS-lasso we use a quadratic B-spline basis, also with six evenly distributed knots. For all methods we use 10-fold cross-validation to find the optimal $\lambda$ for prediction.  

We generate $w_{i1}, ..., w_{iP}, u_i$ and $v_i$ independently from $N(0,1)$ truncated to $[0,1]$. Then, following \citet{Huang2010} the covariates are generated as follows, 
\begin{equation} x_{ij} = \frac{w_{ij}+t u_i}{1+t} \quad \text{for } j \in A, \quad x_{ij} = \frac{w_{ij}+t v_i}{1+t} \quad \text{for } j \not \in A,\label{eq:simdep} \end{equation}
such that if $A$ is the set of components in the true model, the nonzero and zero components are independent, and the correlation among the covariates within the active and nonactive set respectively is controlled by $t$. In all situations the number of replications is 100, $P=1000$, and $n=50$.

To define the true models we consider 
\begin{equation*}
 g^0_1(x) = -\exp(x^2), \quad g^0_2(x) = -\log(x+0.1), \quad
 g^0_3(x) = 2 \tanh(20 x^2) + 0.5 \exp(x^3),
\end{equation*}
\begin{equation*}
 g^0_{4a}(x) = \frac{2\exp(10x-5)}{1+\exp(10x-5)},
\quad g^0_{4b}(x) = 2 x. 
\end{equation*}

Finally, the response variable is generated from the following models; \vspace{2mm}
\begin{itemize}
\item[] Model A $y_i = g^0_1({x}_{i1})+g^0_2({x}_{i2})+g^0_3({x}_{i3})+g^0_{4a}({x}_{i4})+{\epsilon}_i,$\\
\item[] Model B ${y}_i = g^0_1({x}_{i1})+g^0_2({x}_{i2})+g^0_3({x}_{i3})+g^0_{4b}({x}_{i4})+{\epsilon}_i,$
\end{itemize}
\vspace{2mm}
where $\epsilon_i \sim N(0, \sigma^2)$ and $\sigma^2$ is chosen to control the signal to noise ratio ($SNR$). Hence Model A contains only nonlinear monotone effects, while in Model B the fourth component is exchanged with a strictly linear effect. 

Variable selection is compared in terms of the number of experiments for which the method selects each of the true components in the true model, as well as the overall number of true positives (TP) and false positives (FP). We also evaluate the fitted functions for the components in the true model by computing the mean squared error (MSE) between the true function value and the fitted value in the observed points. 

\subsection{Model A} 
For Model A we consider several situations including both independent ($t=0$) and dependent covariates ($t=1$), as well as two typical values for the signal to noise ratio; $SNR \approx 4$ and $SNR \approx 2$. Table \ref{tab:A} reports the results for all six methods in experiments with $SNR \approx 4$ and $t=0$ in expression (\ref{eq:simdep}). For this situation we observe that the MS-lasso is able to select the four components that are really in the true model in nearly all simulation runs, hence the number of true positives is close to four. The MS-lasso has, however, a considerable number of false positives. Introducing an adaptive step reduces the number of false positives and the adaptive MS-lasso is best in recovering the true model. The two methods estimating linear effects, the standard lasso and the adaptive lasso, are also quite good in selecting the true nonzero components, but not as good as the MS-lasso methods and none of them are able to reduce the number of false positives sufficiently. The strongest competitor in this case, is the adaptive liso which performs well in variable selection. However, it is not better than the adaptive MS-lasso in terms of TP and FP. Finally the BS-lasso performs the worst; three of the true nonzero components are selected in less than $25\%$ of the cases. 

\begin{table}
\begin{center}
\caption{Model A: Comparison of the selection and estimation performance for the six methods. To compare the selection performance, the proportion of correct selections in each component in the true model is reported together with the average number of true and false positives. For estimation, MSE between the fitted function and the true function, averaged over the 100 simulated data sets, is reported. }
\label{tab:A}
\footnotesize
\begin{tabular}{lrrrrrr}
  \hline
&&&&&& \\
\multicolumn{7}{c}{Selection} \\
  &&&&&& \\
  & \multicolumn{1}{c}{$g^0_1$} & \multicolumn{1}{c}{$g^0_2$} & \multicolumn{1}{c}{$g^0_3$} & \multicolumn{1}{c}{$g^0_{4a}$} & \multicolumn{1}{c}{TP} & \multicolumn{1}{c}{FP} \\ 
  \hline
  MS-lasso & 1.00 (0.00) & 0.89 (0.31) & 1.00 (0.00) & 1.00 (0.00) & 3.89 (0.31) & 17.72  (9.84) \\ 
  Ad. MS-lasso & 0.98 (0.14) & 0.87 (0.34) & 1.00 (0.00) & 1.00 (0.00) & 3.85 (0.41) & 2.97  (3.05) \\ 
  Lasso & 0.85 (0.36) & 0.72 (0.45) & 1.00 (0.00) & 1.00 (0.00) & 3.57 (0.62) & 25.01(11.85) \\ 
  Ad. lasso & 0.81 (0.39) & 0.68 (0.47) & 1.00 (0.00) & 1.00 (0.00) & 3.49 (0.67) & 18.40  (7.72) \\ 
   Ad. liso & 0.39 (0.49) & 0.98 (0.14) & 1.00 (0.00) & 1.00  (0.00) & 3.37 (0.51) & 5.81  (2.46) \\ 
  BS-lasso & 0.00 (0.00) & 0.04 (0.20) & 0.23 (0.42) & 0.93 (0.26) & 1.20 (0.62) & 1.09  (1.87) \\ 
      \hline
&&&&&& \\
\multicolumn{7}{c}{Estimation} \\
&&&&&& \\
  & \multicolumn{1}{c}{$g^0_1$} & \multicolumn{1}{c}{$g^0_2$} & \multicolumn{1}{c}{$g^0_3$} & \multicolumn{1}{c}{$g^0_{4a}$} & & \\ 
  \hline
  MS-lasso & 0.06 {(0.03)} & 0.17 {(0.08)} & 0.15 {(0.06)} & 0.14 {(0.06)} & \\ 
  Ad. MS-lasso & 0.02 {(0.03)} & 0.07 {(0.10)} & 0.03 {(0.02)} & 0.03 {(0.04)} & \\ 
  Lasso & 0.11 {(0.04)} & 0.26 {(0.06)} & 0.35 {(0.07)} & 0.21 {(0.07)} & \\ 
  Ad. lasso & 0.09 {(0.05)} & 0.22 {(0.09)} & 0.28 {(0.06)} & 0.15 {(0.07)} & \\ 
   Ad. liso & 0.12 {(0.05)} & 0.07 {(0.06)} & 0.08 {(0.04)} & 0.05 {(0.02)} & \\
  BS-lasso & 0.16 {(0.00)} & 0.32 {(0.03)} & 0.62 {(0.22)} & 0.15 {(0.13)} & \\ 
   \hline
\end{tabular}
\end{center}
\end{table}

\begin{figure}
\begin{center}
\subfigure[]{{\includegraphics[width = 0.65\textwidth]{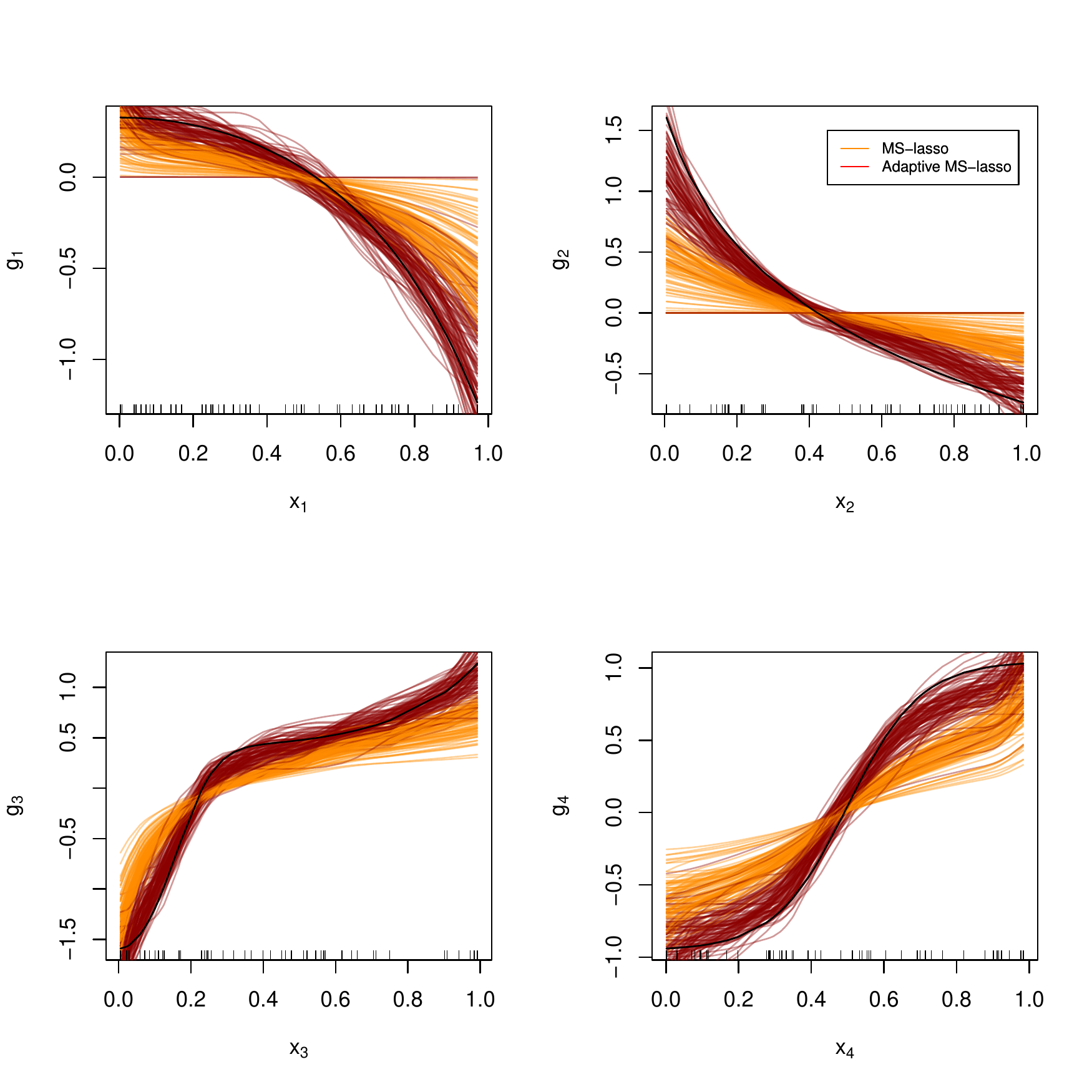}\label{fig:sim:mon}}}
\subfigure[]{{\includegraphics[width = 0.65\textwidth]{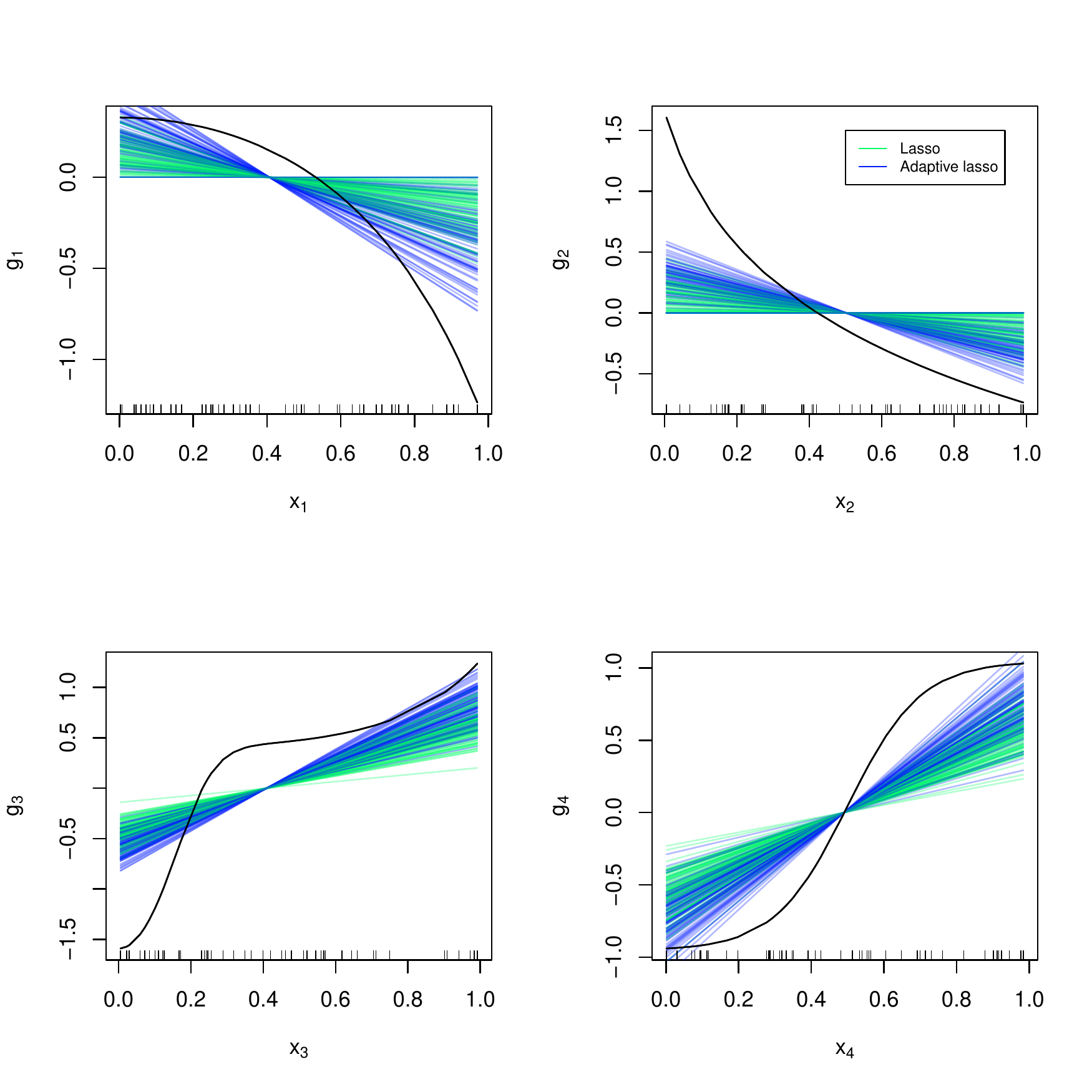}\label{fig:simlin}}}
\caption{Estimated functions for (a) the MS-lasso and the adaptive MS-lasso, and (b) the lasso and adaptive lasso. Each line represents one simulation run. The black line is the true curve. }
\label{fig:simtrans1}
\end{center}
\end{figure}

\begin{figure}
 \centering
\subfigure[]{{\includegraphics[width = 0.65\textwidth]{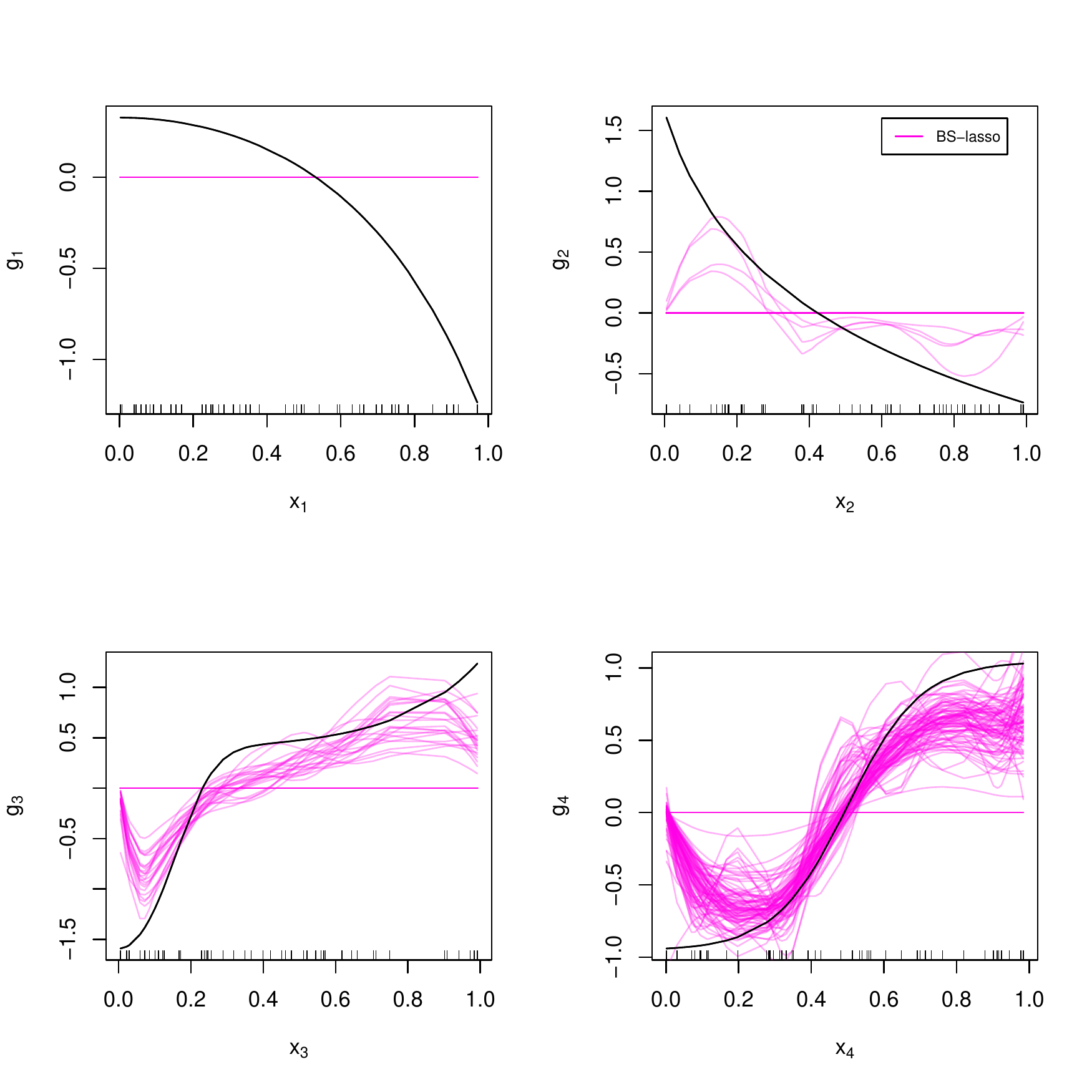}\label{fig:sim:huang}}}
\subfigure[]{{\includegraphics[width = 0.65\textwidth]{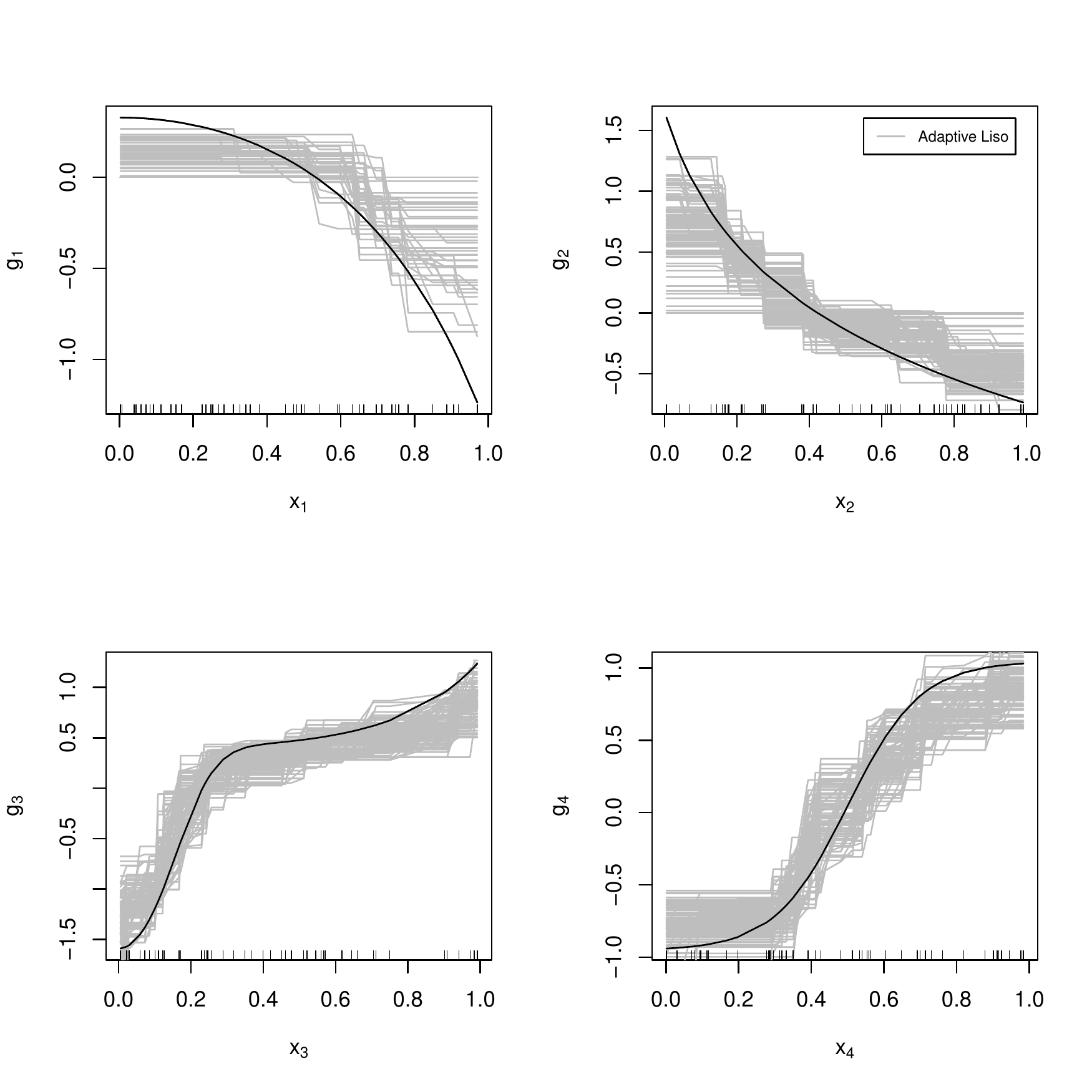}\label{fig:simliso}}}
\caption{Estimated functions for (a) the BS-lasso and (b) the adaptive liso. Each line represents one simulation run. The black line is the true curve. }
\label{fig:simtrans2}
\end{figure}

We also evaluate the estimation error which is given in the lower part of Table \ref{tab:A}. Comparing the MS-lasso and adaptive MS-lasso with their linear competitors, the lasso and the adaptive lasso respectively, we see that the two methods allowing for a nonlinear monotone relationship, estimate the effect of the components with more accuracy. Comparing the MSE for each of the four components individually, we see that the adaptive MS-lasso does much better than all of the other methods. This can also be observed directly from Figure \ref{fig:simtrans1} and \ref{fig:simtrans2} which show the estimated component functions $\hat{g}_j, j = 1, ..., 4$ for all 100 replications. More specifically, the curves for the MS-lasso and the adaptive MS-lasso are plotted in the four upper panels of Figure \ref{fig:simtrans1}, while the curves for the lasso and the adaptive lasso are plotted in the four lower panels. Figure \ref{fig:simtrans2} shows estimated curves for the BS-lasso and the adaptive liso. It is important to note that in Figure \ref{fig:simtrans1}, the nonadaptive and adaptive version of each procedure are plotted in the same panels to illustrate the advantage of introducing an adaptive step. This might give the false impression that there is more variability in the estimated curves, hence it is important to have in mind that in Figure \ref{fig:simtrans1}, the estimated curves come from $2 \times 100$ simulation runs and should be considered separately as indicated by color. Obviously the adaptive MS-lasso captures the true shape of the functions better than the nonadaptive MS-lasso, but also better than all the other methods. Note that the adaptive liso also seems to capture the overall shape of the functions, but the estimation error becomes larger than for the MS-lasso because the estimated functions are step functions and hence not smooth. The BS-lasso achieves large error in all components. Figure \ref{fig:simtrans2} shows that the method is often not able to select the four components, and if they are selected, the estimated effects are heavily oscillating and deviating from the true function. The results of the other experiments conducted for Model A are reported in Appendix A and commented below. 

First we consider the same scenario with dependent covariates ($t=1$). The results are reported in Table 4 in Appendix A. Comparing Table \ref{tab:A} and Table 4, we see that when the covariates are not independent, all methods have more difficulties selecting the correct components. Overall, the MS-lasso performes best in terms of true and false positives. For the fourth component both liso and BS-lasso achieve low MSE, while the MS-lasso and the adaptive MS-lasso performs better or comparable for the three other components. However, for this scenario all methods have problems selecting the first two components, and the adaptive liso and BS-lasso tend to select only the fourth component. In Table 5 in the Supplementary Material we observe that when increasing the sample size to $n=100$ for the dependent covariates, the selection performance is improved for all methods. Our proposed methods are still overall best in selection when considering both true and false positives. As expected we also observe that the MSE's for each component are smaller for all of the methods compared to the results in Table 4. 

The results for the situation with more noise are reported in Table 6 in Appendix A. Obviously the performance of all the methods deteriorates when introducing more noise. The MS-lasso does best in terms of finding the four components in the true model. In terms of estimation, our proposed method is always better or comparable with the linear methods, and always better than B-splines lasso. The adaptive liso is able to select two of the components with highest accuracy, also leading to a smaller MSE for these functions. It seems reasonable that the adaptive liso which provides quite rough estimates of the smooth functions, performs better than the other methods in situations with more noise.

\subsection{Model B}
The results for Model B, with signal to noise ratio $SNR = 4$ and $t=0$ are reported in Table 7 in Appendix A. The true model has three nonlinear monotone effects, while the last component is stricty linear. In this setting we are interested in investigating the performance if there are both nonlinear monotone effects and linear effects in the model. We should be able to select linear effects by using the MS-lasso, as linear functions are special cases of monotone functions. This is confirmed in the results, and in fact the MS-lasso does just as well, if not better, in terms of false positvies as for Model A. Compared to the other methods, the adaptive MS-lasso is overall better in terms of both selection and estimation. 

\subsection{Linear Effects Only}
The MS-lasso allows for nonlinear shapes in the estimated functions, which cannot be captured by the linear methods. However, if the effects are really linear, (adaptive) MS-lasso is still able to estimate the true effects. In these experiments we consider the situation when \textit{all} nonzero effects are linear. We generate $\bm{y}$ from the linear model;
${y}_i = \beta_1{x}_{i1}+\beta_2{x}_{i2}+\beta_3{x}_{i3}+\beta_4{x}_{i4}+{\epsilon}_i,$
with $(\beta_1, \beta_2, \beta_3, \beta_4) = (-2, -2, 2, 2)$. The covariates and the noise are generated as before, with $SNR = 4$. Table \ref{tab:Lin} shows that the (adaptive) MS-lasso is not much worse than the standard lasso and the adaptive lasso which are restricted to linear effects. The linear methods achieve a bit smaller MSE, but in terms of selection, the proposed method is doing just as well. The number of false positives for the MS-lasso and the adaptive MS-lasso is much lower compared to the lasso and the adaptive lasso respectively. The results indicate, however, that both the adaptive liso and B-splines lasso will loose compared to our approach, both in terms of selection and in the ability to fit a straight line. This shows that the possible gain of applying our method is remarkable if the true functions are monotone, but have nonlinear characteristics as in Model A. However, if the effects are truly linear, we are comparable to the methods restricted to linear shapes only. Hence, in general there is much to gain and little to loose using MS-lasso instead of lasso.

\begin{table}
\begin{center}
\caption{All linear: Comparison of the selection and estimation performance for the six methods. To compare the selection performance, the proportion of correct selections for each component in the true model is reported together with the average number of true and false positives. For estimation, the MSE between the fitted function and the true function, averaged over the 100 simulated data sets, is reported. }
\label{tab:Lin}
\small
\footnotesize
\begin{tabular}{lrrrrrr}
  \hline
&&&&&& \\
\multicolumn{7}{c}{Selection} \\
  &&&&&& \\
  & \multicolumn{1}{c}{$g^0_1$} & \multicolumn{1}{c}{$g^0_2$} & \multicolumn{1}{c}{$g^0_3$} & \multicolumn{1}{c}{$g^0_4$} & \multicolumn{1}{c}{TP} & \multicolumn{1}{c}{FP} \\ 
  \hline
  MS-lasso & 1.00 (0.00) & 1.00 (0.00) & 1.00 (0.00) & 1.00 (0.00) & 4.00 (0.00) & 16.26 (8.52) \\ 
  Ad. MS-lasso & 1.00 (0.00) & 1.00 (0.00) & 1.00 (0.00) & 1.00 (0.00) & 4.00 (0.00) & 0.72 (2.29) \\ 
  Lasso & 1.00 (0.00) & 1.00 (0.00) & 1.00 (0.00) & 1.00 (0.00) & 4.00 (0.00) & 27.82 (9.90) \\ 
  Ad.lasso & 1.00 (0.00) & 1.00 (0.00) & 1.00 (0.00) & 1.00 (0.00) & 4.00 (0.00) & 14.14 (4.72) \\ 
  Ad. liso & 0.98 (0.14) & 0.99 (0.10) & 0.98 (0.14) & 0.89 (0.31) & 3.84 (0.55) & 5.80 (2.94) \\ 
  BS-lasso & 0.76 (0.43) & 0.87 (0.34) & 0.43 (0.50) & 0.45 (0.50) & 2.51 (1.38) & 5.13 (4.38) \\ 
\hline
&&&&&& \\
\multicolumn{7}{c}{Estimation} \\
&&&&&& \\
  & \multicolumn{1}{c}{$g^0_1$} & \multicolumn{1}{c}{$g^0_2$} & \multicolumn{1}{c}{$g^0_3$} & \multicolumn{1}{c}{$g^0_4$} & & \\
  \hline
  MS-lasso & 0.02 (0.01) & 0.05 (0.02) & 0.05 (0.02) & 0.07 (0.03) &\\ 
  Ad. MS-lasso & 0.01 (0.00) & 0.04 (0.02) & 0.02 (0.01) & 0.03 (0.02) & \\
  Lasso & 0.02 (0.01) & 0.03 (0.02) & 0.03 (0.02) & 0.04 (0.02) & \\
  Ad. lasso & 0.01 (0.01) & 0.01 (0.01) & 0.01 (0.01) & 0.02 (0.01) & \\ 
  Ad. liso & 0.06 (0.05) & 0.05 (0.04) & 0.07 (0.06) & 0.10 (0.10) & \\
  BS-lasso & 0.16 (0.10) & 0.17 (0.08) & 0.26 (0.10) & 0.23 (0.12) & \\ 
      \hline
\end{tabular}
\end{center}
\end{table}

\section{Data Illustrations}
\label{sec:appl}
In this section we illustrate the use of our method on two relevant data sets from genomics. 

\subsection{Bone Mineral Data}
For illustration we apply our method to a bone mineral data set previously studied in \citet{BoneData}. For 84 women who had a trans-iliacal bone biopsy, there are gene expression measurements for 22815 genes. The data were already normalized as described in \citet{BoneData} and we fit a regression model with bone mineral density as response and gene expressions as covariates. For simplicity and computational reasons in this illustration, we reduce the dimension to $P=2000$ covariates which have the larger empirical variance.

The MS-lasso is used to model the relationship between the bone mineral density and the expression of the genes. The covariates are transformed to $[0,1]$ and we use I-splines of order two with six interior knots. We compare the results with; the lasso and the adaptive lasso for linear effects, the adaptive liso for isotonic effects, as well as the BS-lasso for nonparametric additive models \citep{Huang2010}. For the latter, quadratic B-splines with six evenly distributed knots are used. For all methods we use 10-fold cross-validation to choose the penalty parameters. 

The six methods select, as expected, different sets of genes. The MS-lasso selects 19 genes, while when including an adaptive step, the adaptive MS-lasso selects 8 genes. The lasso and the adaptive lasso select 14 and 12 variables respectively, while the adaptive liso selects 13 variables. Finally, the BS-lasso selects 8 variables. Out of the 19 genes selected by the MS-lasso, the lasso selects 10 of them and the adaptive lasso 9 of the same genes. The adaptive liso finds 7 of the same genes as the MS-lasso and the BS-lasso finds 5 of the same.

\begin{figure}
\centering
\subfigure[]{\includegraphics[width = 0.41\textwidth]{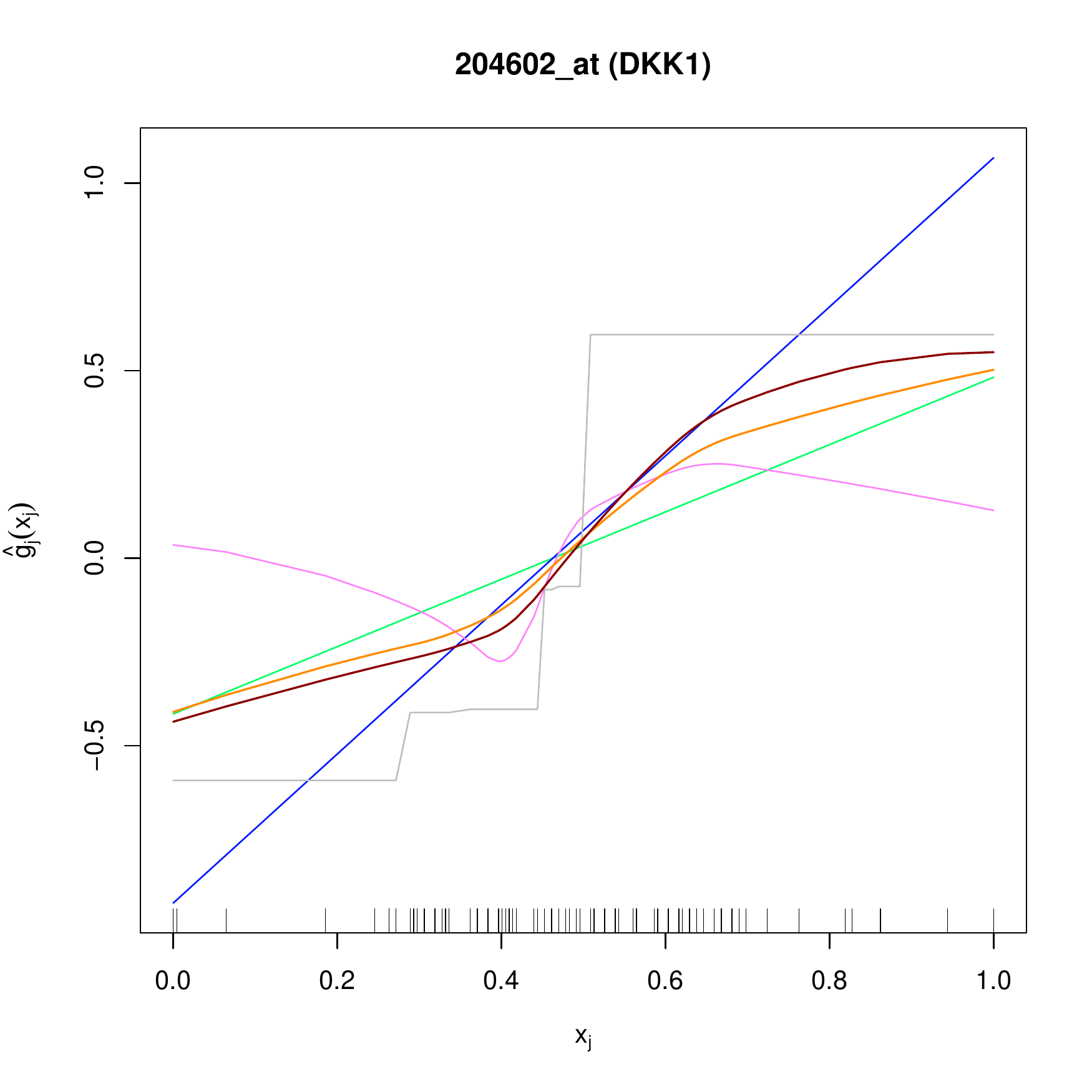}\label{fig:bone1}}
\subfigure[]{\includegraphics[width = 0.41\textwidth]{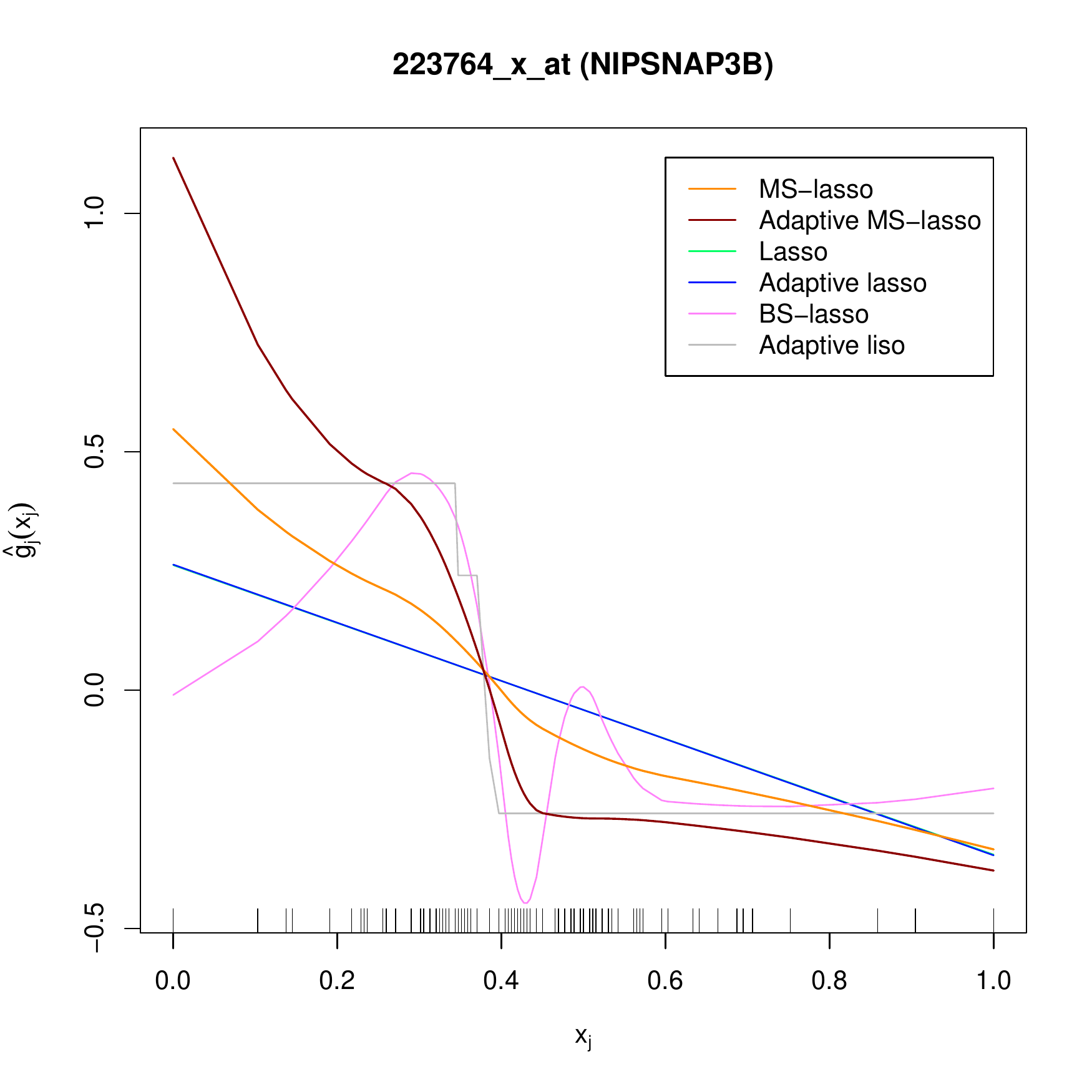}\label{fig:bone2}}
\subfigure[]{\includegraphics[width = 0.41\textwidth]{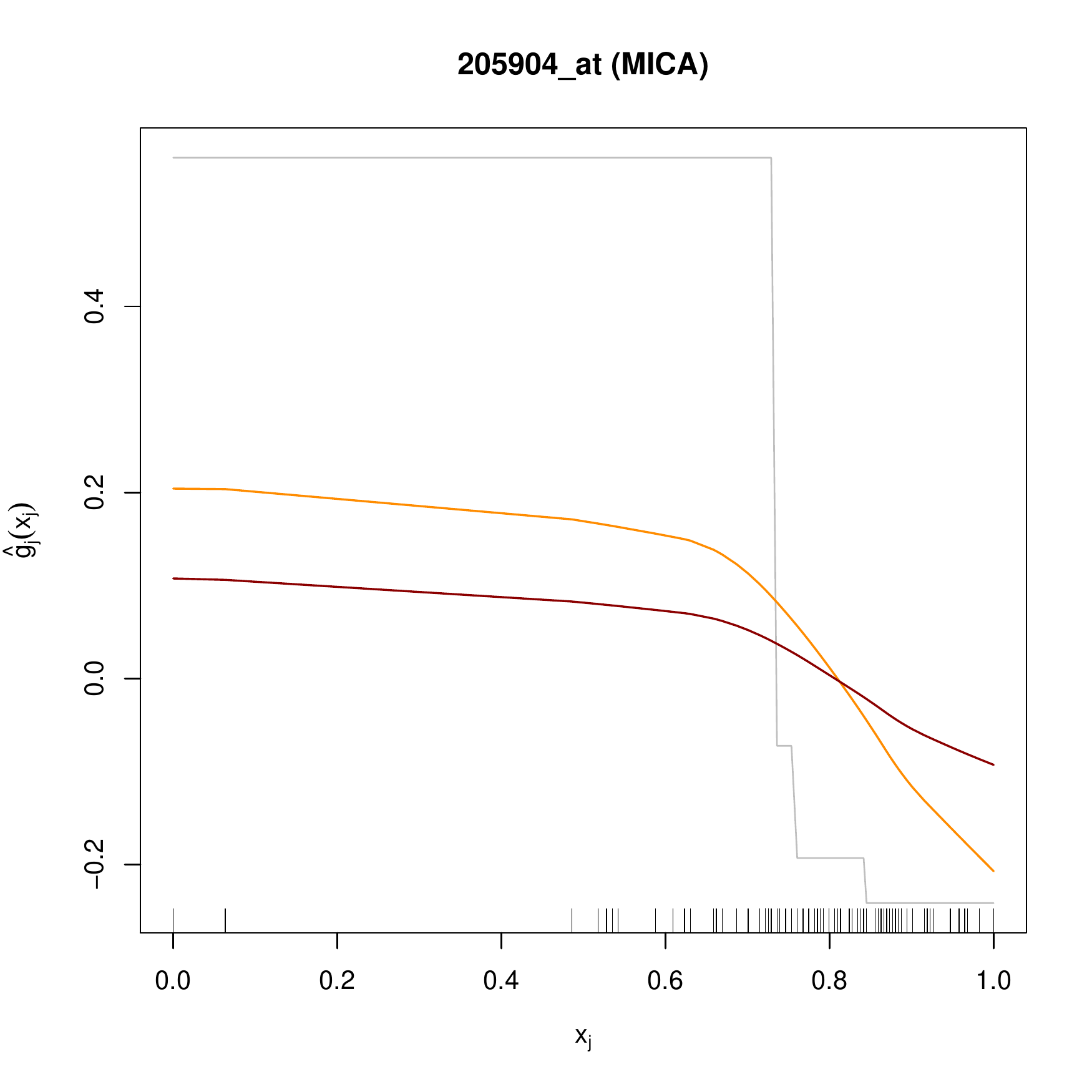}\label{fig:bone3}}
\subfigure[]{\includegraphics[width = 0.41\textwidth]{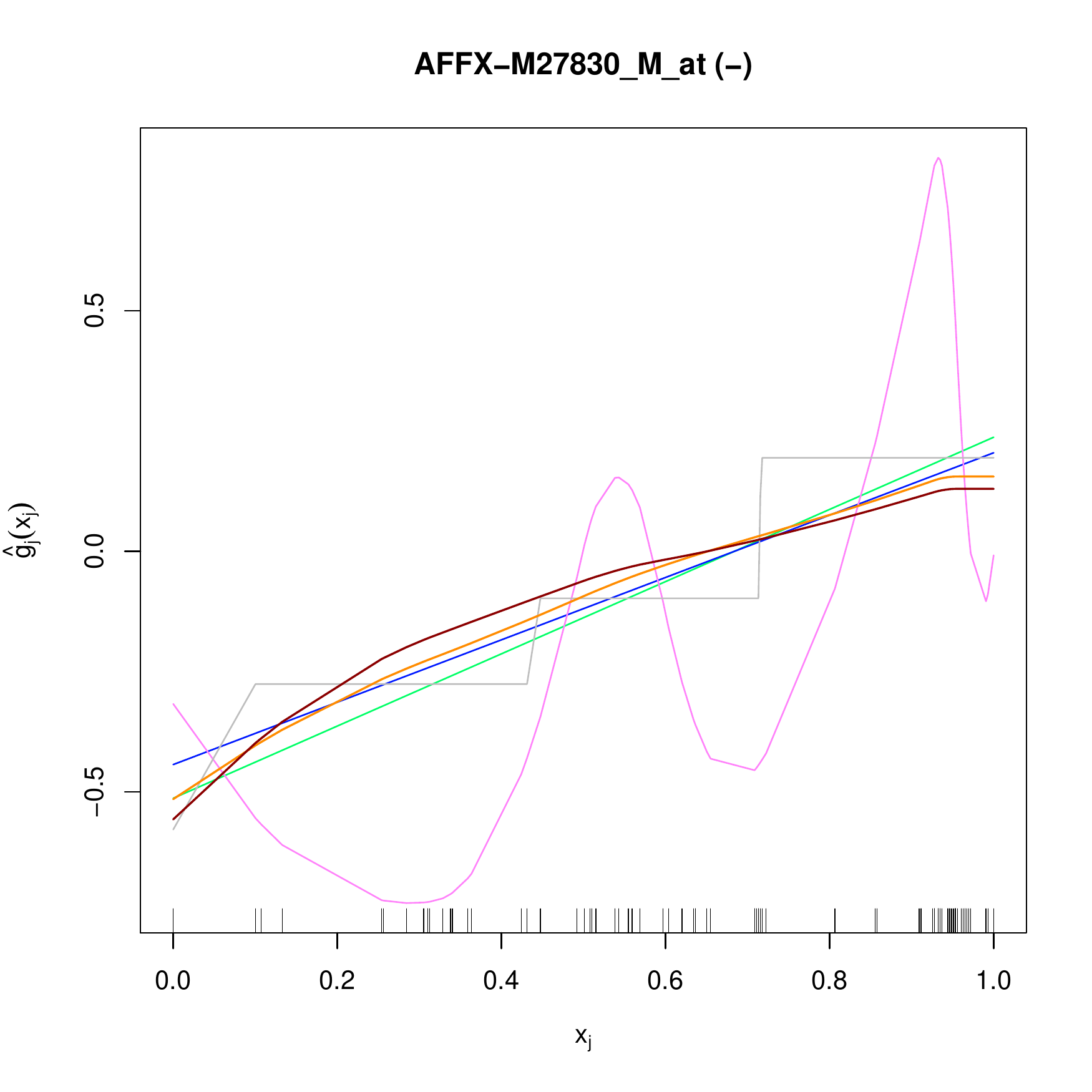}\label{fig:bone4}}
\caption{Estimated functions for four selected genes from the bone data example. }
\label{fig:bone2000}
\end{figure}
In Figure \ref{fig:bone2000}, we have plotted the estimated functions for four of the variables that are selected by our method together with the estimated effects by the other methods, if the variable is selected by any of them. Figure \ref{fig:bone1} shows the estimated effects for a variable which is selected by all of the methods. The three methods assuming monotonicity indicate a slow increase for low values of the gene expression, a more rapid increase for midrange values and a plateau effect for high expression values. Obviously, the linear methods are not able to capture this shape, and the BS-lasso is estimating a clear nonmonotone shape which is less reasonable in the current biological applicaton. 

In Figure \ref{fig:bone2} we see a similar pattern where the adaptive MS-lasso and the adaptive liso recognize a rapid decrease for midrange values, and it seems that there might be a threshold effect. BS-lasso also seems to be recognizing this rapid decrease, but again the method seems too flexible. In Figure \ref{fig:bone3} we show an example where only the methods assuming monotonicity are selecting the component. Both the monotone splines methods and the adaptive liso estimate a decreasing effect with a breakpoint around 0.7 (in the transformed values). The effect beyond 0.7 seems to be quite linear, but since the linear methods are not able to capture the threshold effect, they are not selecting the component. The final plot in Figure \ref{fig:bone4} illustrates an effect that most likely is linear. We observe that our methods seem to be able to capture the linearity quite well, with an estimated effect that is close to linear. Again we see that BS-lasso seems to suffer from too much flexibility, with a very wiggly and nonlinear estimate of the component effect that is less likely in this situation.

\subsection{Breast cancer data}
\label{sec:appl2}
In cancer studies, recovering relationships between copy number variations (CNV) and gene expression data is of interest as such biological relationships can be relevant for the understanding of disease. Many methods that have been used to investigate these relationships assume linearity, while the underlying biological interaction is not really believed to be linear \citep{Hiroko2011}. \citet{Hiroko2011} employed a strategy allowing for non-linear dependencies when studying in-cis how a copy number variation would impact the expression. In this example we propose to use MS-lasso to study the dependency between copy number variation and expression also in-trans, assuming monotone nonlinear relationships. Our approach can be used to investigate which CNV's impact in-trans the expression of a gene G. By defining the response $\bm{y}$ as the gene expression of gene G and considering the CNV's as covariates, the aim is to select a set of CNV which describes the variation in gene expression for gene G. 

For illustration, we use the combined copy number - gene expression data studied in \citet{Hiroko2011} and consider one of the genes, ERBB2, which is discussed in their paper as target variable in the analysis. That is, the gene expression measurements are considered as the response while the copy number variations within the same chromosome as ERBB2 are used as covariates. This becomes a regression problem with $n=102$ and $P=668$ covariates. 

\begin{figure}
\centering
\subfigure[]{\includegraphics[width = 0.45\textwidth]{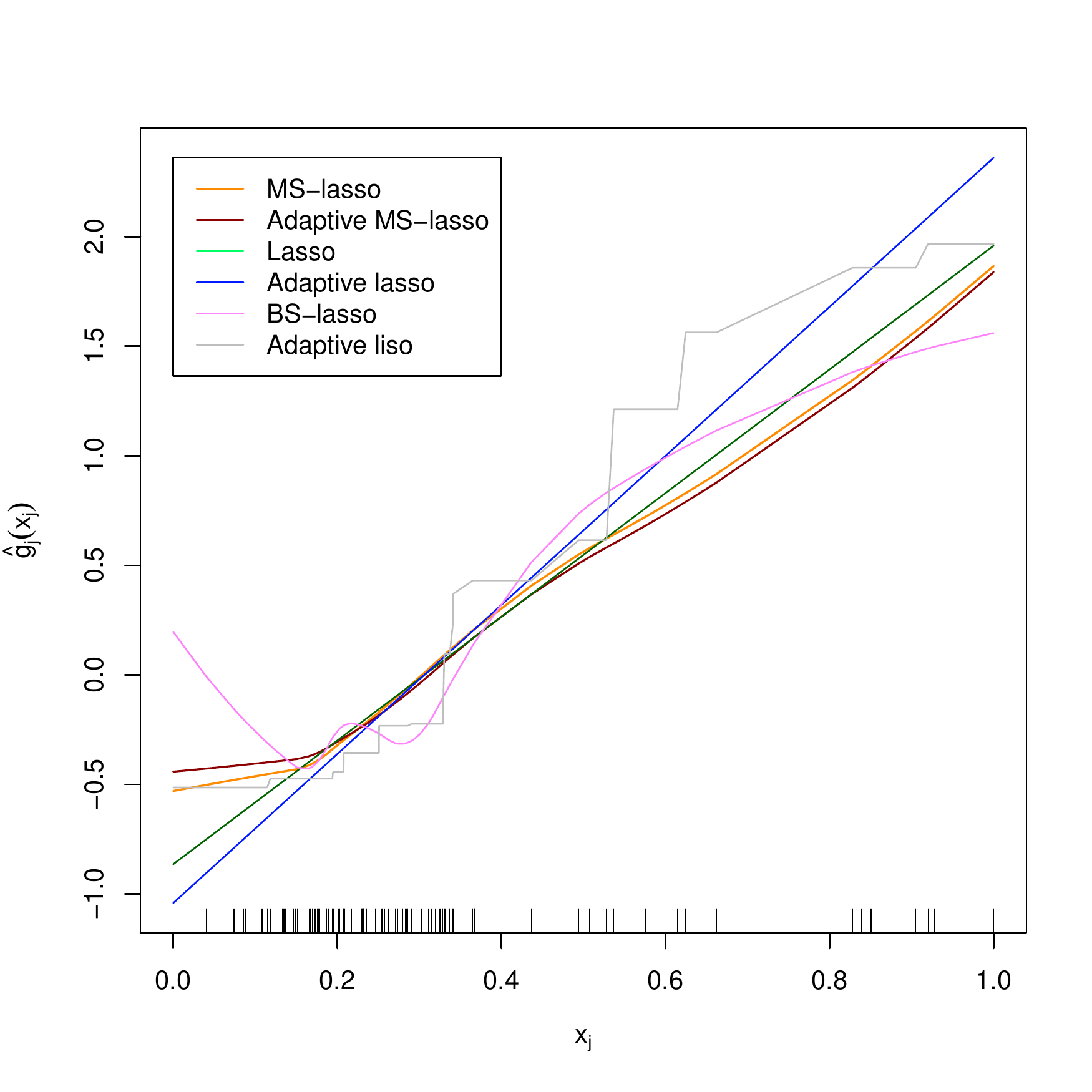}\label{fig:breastcis}}
\subfigure[]{\includegraphics[width = 0.45\textwidth]{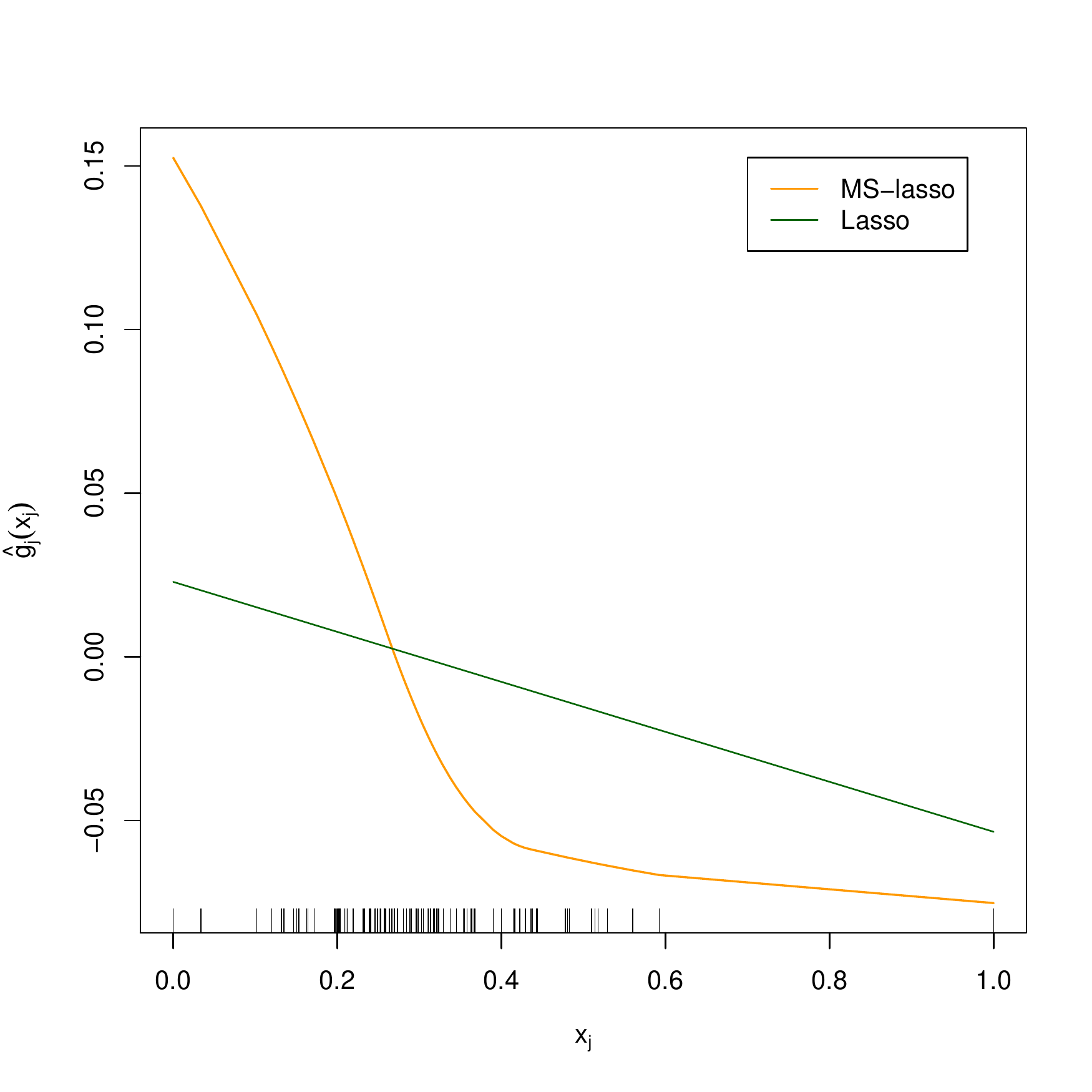}\label{fig:breast}}
\caption{Estimated functions for two selected CNVs from the breast cancer illustration. }
\label{fig:breastfull}
\end{figure}
In this case, the MS-lasso selects six CNV's. Interestingly, one of them is the copy number corresponding to the gene ERBB2 itself, which is in fact selected by all of the methods. The estimated function is plotted in Figure \ref{fig:breastcis}. The plot indicates that this effect is actually quite linear, as all methods seem to estimate a shape that is close to linear. However, we also here observe that the estimated linear effects might not capture fully the shapes. For example in Figure \ref{fig:breast} we compare the estimated effects of the MS-lasso and the lasso for another of the selected CNV. From the function estimated by our method, it seems to be a steep decrease with a threshold around 0.4 (in the transformed values of the covariate). The linear estimate is not capturing such a shape and seems to be highly influenced by the few large values in the covariate. The other methods have not selected this CNV.

\section{Discussion}
\label{sec:disc}
The additive model provides a flexible alternative to the standard linear regression model. However, the monotonicity in the linear model is attractive and in some settings it is sensible to preserve monotonicity in the additive models by imposing restrictions on the additive components. In this paper we have proposed a new method for estimation and variable selection in high-dimensional additive models that is restricted to monotone effects. By combining group selection with spline representations of the covariates, the method is within the same category as the methods of for example \citet{Buhlmann2009} and  \citet{Huang2010} allowing for nonlinearity in the estimated functions. 

We have presented the MS-lasso as a novel and flexible method feasible in high-dimensional settings when $P>n$. Also in low dimensions, if the covariates are represented by a larger number of basis functions or in any other regression setting where variable selection is relevant, our MS-lasso could be useful. The MS-lasso can also be extended to generalized additive models. In this case we simply replace the residual sum of squares in (\ref{beta-mono}) with the relevant (partial) log-likelihood, still representing the covariates by their I-splines basis. 

Similar to \citet{Buhlmann2009} or \citet{Huang2010} among others, we fit smooth curves with a fixed number of knots (typically six interior knots), although the placement and number of knots are important issues within the splines literature. \citet{Meyer2008} argues that assumptions on the shape of the function, such as monotonicity, are quite robust to the choice of knots and therefore avoids much of the problems of overfitting the data which are present using other types of (regression) splines. Also, \citet{smoothingregression} argue that the number of knots recommended using monotone I-splines is usually much smaller than the number of observation points. Following these arguments, we have not studied the choice of knots in further detail, but this could be a topic of future research.

Another important issue is related to computational efficiency. Though the MS-lasso is feasible when $P>n$, the computational complexity may increase dramatically depending on the number of knots. In situations where the number of inputs to the algorithm becomes too large, it could be reasonable to employ suitable screening or preselection techniques to reduce computational complexity.

We should also address a minor detail about the choice of the penalty parameter $\lambda$. As pointed out in \citet{coop}, when the cooperative penalty goes to zero, that is for small $\lambda$, sign-coherence is no longer preserved for the cooperative lasso. Hence monotonicity is neither preserved if $\lambda$ is too small. This rarely occurs in our experiments, but to ensure monotonicity we should consider only values of $\lambda$ for which sign-coherence of the cooperative lasso is preserved. 

Recently, several approaches doing variable selection in additive partial linear models have been proposed. For example do \citet{Carrol2011} and \citet{Liu2011} introduce estimation and variable selection for additive partial linear models with a predefined division of covariates into linear and nonlinear associations with the response. Similarly one may construct models with some covariates strictly linearly related to the response, while a second part of the covariates have nonlinear monotone effect on the response. This is simply achieved by letting the group size in the MS-lasso be equal to one for the linear part of the covariates, indicating that it is straightforward to extend the MS-lasso also to additive partial linear models.

Based on our simulation results and data examples, we conclude that the (adaptive) MS-lasso shows excellent properties compared to other methods and should be advised for high-dimensional monotone regression.

\section*{Acknowledgements}
We thank Hiroko Solvang for providing the breast cancer data, and Sjur Reppe who provided the bone biopsy data. \vspace{-0.3cm}

\appendix
\section*{Appendix A}
\markboth{L.C. Bergersen, K. Tharmaratnam, and I.K. Glad}{Monotone Splines Lasso}
This section contains four tables with results from simulation experiments. The first two tables, Table 4 and Table 5 report the results from experiments using Model A, as described in the paper, and dependent covariates. We consider sample size $n=50$ in Table 4, and $n=100$ in Table 5. In Table 6, results from Model A with more noise are reported. Finally, Table 7 contains the results from Model B. All experiments are described and discussed in Section 3. 
\begin{table}
\centering
\caption{Results for Model A with $t=1$ (dependent covariates) and $SNR = 4$: Comparison of the selection and estimation performance for the six methods. To compare the selection performance, the proportion of which the methods select each component in the true model is reported together with the average number of true and false positives. For estimation the mean squared error between the fitted function and the true function, averaged over the 100 simulated data sets, is reported. }
\small
\footnotesize
\begin{tabular}{lrrrrrr}
  \hline
&&&&&& \\
\multicolumn{7}{c}{Selection} \\
  &&&&&& \\
  & \multicolumn{1}{c}{$g^0_1$} & \multicolumn{1}{c}{$g^0_2$} & \multicolumn{1}{c}{$g^0_3$} & \multicolumn{1}{c}{$g^0_{4a}$} & \multicolumn{1}{c}{TP} & \multicolumn{1}{c}{FP} \\ 
  \hline
  MS-lasso & 0.16 (0.37) & 0.12 (0.33) & 0.97 (0.17) & 1.00 (0.00)  & 2.25 (0.70) & 4.48 (7.25) \\ 
  Ad. MS-lasso & 0.12 (0.33) & 0.11 (0.31) & 0.71 (0.46) & 0.98 (0.14) & 1.92 (0.85) & 0.20 (0.77) \\ 
  Lasso & 0.00 (0.00) & 0.05 (0.22) & 0.98 (0.14) & 1.00 (0.00) & 2.03 (0.26) & 7.64 (4.01) \\ 
  Ad. lasso & 0.00 (0.00) & 0.05 (0.22) & 0.89 (0.31) & 1.00 (0.00)  & 1.94 (0.40) & 5.87 (3.30) \\ 
  Ad. liso & 0.00 (0.00) & 0.00 (0.00) & 0.04 (0.20) & 1.00 (0.00)  & 1.04 (0.20) & 7.10 (4.83) \\   
  BS-lasso & 0.00 (0.00) & 0.00 (0.00) & 0.00 (0.00) & 1.00 (0.00) & 1.00 (0.00) & 0.82 (1.71) \\ 
   \hline
&&&&&& \\
\multicolumn{7}{c}{Estimation} \\
&&&&&& \\
  & \multicolumn{1}{c}{$g^0_1$} & \multicolumn{1}{c}{$g^0_2$} & \multicolumn{1}{c}{$g^0_3$} & \multicolumn{1}{c}{$g^0_{4a}$} & &\\
  \hline
  MS-lasso & 0.06 (0.01) & 0.16 (0.04) & 0.13 (0.04) & 0.20 (0.05) & \\
  Ad. MS-lasso & 0.06 (0.02) & 0.15 (0.05) & 0.13 (0.07) & 0.16 (0.08) & \\
  Lasso & 0.06 (0.00) & 0.17 (0.01)& 0.16 (0.02) & 0.19 (0.03) & \\
  Ad. lasso & 0.06 (0.00) & 0.16 (0.03) & 0.14 (0.03) & 0.16 (0.04) & \\
  Ad. liso & 0.06 (0.00) & 0.17 (0.00) & 0.21 (0.01) & 0.14 (0.05) & \\
  BS-lasso & 0.06 (0.00) & 0.17 (0.00) & 0.21 (0.00) & 0.12 (0.04) & \\
   \hline
\end{tabular}
\end{table}
\begin{table}
\centering
\caption{Results for Model A with $t=1$ (dependent covariates), $SNR = 4$ and $n = 100$: Comparison of the selection and estimation performance for the six methods. To compare the selection performance, the proportion of which the methods select each component in the true model is reported together with the average number of true and false positives. For estimation the mean squared error between the fitted function and the true function, averaged over the 100 simulated data sets, is reported. }
\small
\footnotesize
\begin{tabular}{lrrrrrr}
  \hline
&&&&&& \\
\multicolumn{7}{c}{Selection} \\
  &&&&&& \\
  & \multicolumn{1}{c}{$g^0_1$} & \multicolumn{1}{c}{$g^0_2$} & \multicolumn{1}{c}{$g^0_3$} & \multicolumn{1}{c}{$g^0_{4a}$} & \multicolumn{1}{c}{TP} & \multicolumn{1}{c}{FP} \\ 
  \hline
  MS-lasso & 1.00 (0.00) & 1.00 (0.00) & 1.00 (0.00) & 1.00 (0.00) & 4.00 (0.00) & 1.53 (1.41) \\ 
  Ad. MS-lasso & 0.78 (0.42) & 1.00 (0.00) & 1.00 (0.00) & 1.00 (0.00) & 3.78 (0.42) & 0.00 (0.00) \\ 
  Lasso & 1.00 (0.00) & 1.00 (0.00) & 1.00 (0.00) & 1.00 (0.00) & 4.00 (0.00) & 23.16 (16.95) \\ 
  Ad. lasso & 1.00 (0.00) & 1.00 (0.00) & 1.00 (0.00) & 1.00 (0.00) & 4.00 (0.00) & 14.68 (8.77) \\ 
  Ad. liso & 0.71 (0.46) & 1.00 (0.00) & 1.00 (0.00) & 1.00 (0.00) & 3.71 (0.46) & 3.27 (2.53) \\ 
  BS-lasso & 0.14 (0.35) & 0.99 (0.10) & 1.00 (0.00) & 1.00 (0.00) & 3.13 (0.37) & 5.81 (5.09) \\ 
   \hline
&&&&&& \\
\multicolumn{7}{c}{Estimation} \\
&&&&&& \\
  & \multicolumn{1}{c}{$g^0_1$} & \multicolumn{1}{c}{$g^0_2$} & \multicolumn{1}{c}{$g^0_3$} & \multicolumn{1}{c}{$g^0_{4a}$} & &\\
  \hline
  MS-lasso & 0.02 (0.01) & 0.03 (0.01) & 0.02 (0.00) & 0.05 (0.01) & \\
  Ad. MS-lasso & 0.04 (0.02) & 0.02 (0.01) & 0.01 (0.00) & 0.06 (0.03) & \\
  Lasso & 0.02 (0.01) & 0.03 (0.01) & 0.06 (0.00) & 0.05 (0.01) & \\
  Ad. lasso & 0.02 (0.00) & 0.02 (0.01) & 0.05 (0.00) & 0.03 (0.01) & \\
  Ad. liso & 0.04 (0.02) & 0.01 (0.00) & 0.03 (0.01) & 0.03 (0.01) & \\
  BS-lasso & 0.07 (0.02) & 0.02 (0.02) & 0.03 (0.02) & 0.04 (0.02) & \\
      \hline
\end{tabular}
\end{table}

\begin{table}
\centering
\caption{Results for Model A with $SNR = 2$: Comparison of the selection and estimation performance for the six methods. To compare the selection performance, the proportion of which the methods select each component in the true model is reported together with the average number of true and false positives. For estimation the mean squared error between the fitted function and the true function, averaged over the 100 simulated data sets, is reported. }
\small
\footnotesize
\begin{tabular}{lrrrrrr}
  \hline
&&&&&& \\
\multicolumn{7}{c}{Selection} \\
  &&&&&& \\
  & \multicolumn{1}{c}{$g^0_1$} & \multicolumn{1}{c}{$g^0_2$} & \multicolumn{1}{c}{$g^0_3$} & \multicolumn{1}{c}{$g^0_{4a}$} & \multicolumn{1}{c}{TP} & \multicolumn{1}{c}{FP} \\ 
  \hline
  MS-lasso & 0.50 (0.50) & 0.25 (0.44) & 0.93 (0.26) & 0.89 (0.31) & 2.57 (1.04) & 11.74 (9.92) \\ 
  Ad. MS-lasso & 0.33 (0.47) & 0.18 (0.39) & 0.92 (0.27) & 0.73 (0.45) & 2.16 (1.05) & 3.64 (4.27) \\ 
  Lasso & 0.37 (0.49) & 0.25 (0.44) & 0.85 (0.36) & 0.86 (0.35) & 2.33 (1.12) & 19.54 (14.19) \\ 
  Ad. lasso & 0.30 (0.46) & 0.20 (0.40) & 0.85 (0.36) & 0.83 (0.38) & 2.18 (1.05) & 15.15 (10.43) \\ 
  Ad. liso & 0.11 (0.31) & 0.45 (0.50) & 0.83 (0.38) & 0.92 (0.27) & 2.31 (1.03) & 7.73 (4.87) \\ 
  BS-lasso & 0.01 (0.10) & 0.02 (0.14) & 0.13 (0.34) & 0.57 (0.50) & 0.73 (0.69) & 1.57 (2.68) \\ 
   \hline
&&&&&& \\
\multicolumn{7}{c}{Estimation} \\
&&&&&& \\
  & \multicolumn{1}{c}{$g^0_1$} & \multicolumn{1}{c}{$g^0_2$} & \multicolumn{1}{c}{$g^0_3$} & \multicolumn{1}{c}{$g^0_{4a}$} & &\\
  \hline
  MS-lasso & 0.13 (0.05) & 0.30 (0.05) & 0.35 (0.15) & 0.34 (0.15) & \\
  Ad. MS-lasso & 0.12 (0.06) & 0.29 (0.09) & 0.18 (0.18) & 0.27 (0.21) & \\
  Lasso & 0.14 (0.03) & 0.30 (0.05) & 0.48 (0.15) & 0.34 (0.16) & \\
  Ad. lasso & 0.13 (0.05) & 0.29 (0.07) & 0.40 (0.18) & 0.27 (0.19) & \\
  Ad. liso & 0.15 (0.03) & 0.24 (0.11)& 0.28 (0.24) & 0.16 (0.16) & \\
  BS-lasso & 0.16 (0.01) & 0.32 (0.02) & 0.69 (0.14) & 0.35 (0.23) & \\
      \hline
\end{tabular}
\end{table}

\begin{table}
\centering
\caption{Results for Model B with $SNR = 4$: Comparison of the selection and estimation performance for the six methods. To compare the selection performance, the proportion of which the methods select each component in the true model is reported together with the average number of true and false positives. For estimation the mean squared error between the fitted function and the true function, averaged over the 100 simulated data sets, is reported. }
\small
\footnotesize
\begin{tabular}{lrrrrrr}
  \hline
&&&&&& \\
\multicolumn{7}{c}{Selection} \\
  &&&&&& \\
  & \multicolumn{1}{c}{$g^0_1$} & \multicolumn{1}{c}{$g^0_2$} & \multicolumn{1}{c}{$g^0_3$} & \multicolumn{1}{c}{$g^0_{4b}$} & \multicolumn{1}{c}{TP} & \multicolumn{1}{c}{FP} \\ 
  \hline
  MS-lasso & 1.00 (0.00) & 1.00 (0.00) & 1.00 (0.00) & 1.00 (0.00) & 4.00 (0.00) & 22.27 (8.64) \\ 
  Ad. MS-lasso & 0.97 (0.17) & 0.98 (0.14) & 1.00 (0.00) & 0.98 (0.14) & 3.93 (0.43) & 1.41 (2.22) \\ 
  Lasso & 0.93 (0.26) & 0.96 (0.20) & 1.00 (0.00) & 1.00 (0.00) &  3.89 (0.40) & 31.33 (11.32) \\ 
  Ad. lasso & 0.85 (0.36) & 0.93 (0.26) & 1.00 (0.00) & 1.00 (0.00) & 3.78 (0.52) & 21.39 (6.87) \\
  Ad. liso & 0.35 (0.48) & 0.95 (0.22) & 1.00 (0.00) & 0.91 (0.29) & 3.21 (0.67) & 6.95 (3.33) \\  
  BS-lasso & 0.01 (0.10) & 0.08 (0.27) & 0.55 (0.50) & 0.15 (0.36) & 0.79 (0.86) & 3.77 (4.13) \\ 
\hline
&&&&&& \\
\multicolumn{7}{c}{Estimation} \\
&&&&&& \\
  & \multicolumn{1}{c}{$g^0_1$} & \multicolumn{1}{c}{$g^0_2$} & \multicolumn{1}{c}{$g^0_3$} & \multicolumn{1}{c}{$g^0_{4b}$} & &\\
  \hline
MS-lasso & 0.04 (0.02) & 0.10 (0.03) & 0.10 (0.03) & 0.06 (0.03) & \\
  Ad. MS-lasso & 0.02 (0.03) & 0.04 (0.05) & 0.03 (0.04) & 0.02 (0.05) & \\
  Lasso & 0.10 (0.03) & 0.19 (0.06) & 0.27 (0.06) & 0.11 (0.06) & \\
  Ad. lasso & 0.09 (0.04) & 0.15 (0.07) & 0.23 (0.04) & 0.08 (0.06) & \\
  Ad. liso & 0.13 (0.05) & 0.09 (0.07) & 0.09 (0.05) & 0.11 (0.09) & \\
  BS-lasso & 0.16 (0.01) & 0.32 (0.03) & 0.52 (0.22) & 0.31 (0.06) & \\
      \hline
\end{tabular}
\end{table}
\newpage
\bibliographystyle{apalike}
\bibliography{Reference}
\end{document}